\documentclass[journal,10pt]{IEEEtran}
\ifCLASSINFOpdf

\else

\fi

\usepackage[utf8]{inputenc}
\usepackage{amssymb}
\usepackage{stfloats}
\usepackage{placeins}
\usepackage{overpic}
\usepackage[font=small]{caption}
\usepackage{amsmath}
\usepackage{comment}
\usepackage{mathtools}
\usepackage{booktabs}

\usepackage{cite}
\usepackage{subfig}
\usepackage{lipsum}
\usepackage{graphicx}
\usepackage{multicol}
\usepackage{algorithm}
\usepackage[noend]{algpseudocode}
\usepackage{tikz}
\usepackage[english]{babel}
\usepackage{amsthm}
\theoremstyle{plain}
\usepackage{dirtytalk}

\graphicspath{ {./images/} }
\usepackage{amsmath}

\usepackage{mathtools}

\usepackage{tabularx}
\newcolumntype{L}[1]{>{\raggedright\arraybackslash}p{#1}}
\newcolumntype{C}[1]{>{\centering\arraybackslash}p{#1}}
\newcolumntype{R}[1]{>{\raggedleft\arraybackslash}p{#1}}
\usepackage{multirow}
\usepackage{mathtools, cuted}

\begin{document}


\title{Rate-Splitting assisted Massive Machine-Type Communications in Cell-Free Massive MIMO}


\author{\IEEEauthorblockN{Anup~Mishra,  Yijie~Mao, \IEEEmembership{Member, IEEE},  Luca~Sanguinetti, \IEEEmembership{Senior Member, IEEE} \\and Bruno~Clerckx, \IEEEmembership{Fellow, IEEE}\vspace{-0.3cm}}
\thanks{The authors Anup Mishra and Bruno Clerckx are with the Department of Electrical and Electronic Engineering, Imperial College London, London SW7 2AZ,
U.K. (e-mail: anup.mishra17@imperial.ac.uk; b.clerckx@imperial.ac.uk).}
\thanks{Yijie Mao is with the School of Information Science and Technology, ShanghaiTech University, Shanghai 201210, China (e-mail: maoyj@shanghaitech.edu.cn).}
\thanks{Luca Sanguinetti is with the Dipartimento di Ingegneria dell’Informazione,
University of Pisa, 65122 Pisa, Italy (e-mail: luca.sanguinetti@unipi.it).}}


\maketitle


\begin{abstract}
 This letter focuses on integrating rate-splitting multiple-access (RSMA) with time-division-duplex Cell-free Massive MIMO (multiple-input multiple-output) for massive machine-type communications. Due to the large number of devices, their sporadic access behaviour and limited coherence interval, we assume a random access strategy with all active devices utilizing the same pilot for uplink channel estimation. This gives rise to a highly pilot-contaminated scenario, which inevitably deteriorates channel estimates. Motivated by the robustness of RSMA towards imperfect channel state information, we propose a novel RSMA-assisted downlink transmission framework for cell-free massive MIMO. On the basis of the downlink achievable spectral efficiency of the common and private streams, we devise a heuristic common precoder design and propose a novel max-min power control method for the proposed RSMA-assisted scheme. Numerical results show that RSMA effectively mitigates the effect of pilot contamination in the downlink and achieves a significant performance gain over a conventional cell-free massive MIMO network.   
\end{abstract}




\section{Introduction}\label{Intro}


\IEEEPARstart{M}{}assive Machine Type communications (mMTC) is one of the most prominent features of B$5$G and $6$G wireless networks \cite{Schober_mMTC}. The main characteristics of mMTC include a large number of low-power user equipments (UEs), widespread coverage, sporadic traffic and low-data rate requirements~\cite{Schober_mMTC,erykMTC@2017}. To ensure widespread coverage, Cell-free Massive MIMO (CF-MaMIMO) is seen as a promising technology for mMTC~\cite{luca@pilot}. The large number of UEs and their sporadic access behaviour in mMTC may give rise to high pilot contamination in the uplink (UL) training phase~\cite{erykMTC@2017,carvalho@randomaccess}. In fact, the short coherence interval limits the number of orthogonal pilots~\cite{CFMaMIMOVsSmallCells}, whereas sporadic activity makes orthogonal scheduling of UEs infeasible~\cite{carvalho@randomaccess}. To overcome this latter problem, random access techniques are typically employed where active UEs randomly pick a pilot sequence from a small pool of orthogonal sequences~\cite{carvalho@randomaccess}. In mMTC, the probability of multiple UEs sharing the same pilot is very high~\cite{carvalho@randomaccess,luca@pilot}. In \cite{Yin@singlepilot}, the authors consider the worst case in which the same pilot is used by all UEs and study the performance in the UL of CF-MaMIMO. It is well known that pilot contamination deteriorate the channel estimation quality and results into statistically dependent channel estimates~\cite{massivemimobook, luca@pilot, CFMaMIMOVsSmallCells}. Consequently, designing precoders based on such imperfect channel state information (CSI) at the transmitter imposes severe multi-user interference in the downlink (DL), and the network may become strongly interference-limited. 

To mitigate the interference in wireless networks, a powerful interference management strategy, named rate splitting multiple access (RSMA), was introduced in \cite{mao2017rate} and is considered as a promising paradigm for the physical-layer transmission in $6$G \cite{Onur@6G}. In its single-layer form, RSMA splits the message of each UE into a common part and a private part
\cite{mishra2021ratesplitting,mao2017rate}. The common parts of all UE messages are combined together to form a common message that will be decoded by all UEs.
The private part of each UE message is independently encoded and will be decoded by the corresponding UE only. The common message is superimposed on top of the private messages for transmission. By adjusting the message split and power allocated to the common and private messages, RSMA allows to better manage interference. RSMA is known to be robust to imperfect CSI and to outperform conventional linear precoding schemes in conventional multi-user MIMO networks in terms of spectral efficiency (SE) \cite{mishra2021ratesplitting,mao2017rate} and energy efficiency \cite{mao2018EE}. RSMA was also applied to Massive MIMO in~\cite{massiveMIMO@hardware,Minbo2016MassiveMIMO}.

Building on \cite{Yin@singlepilot}, in this letter we consider mMTC with random access and assume that all active UEs utilize the same pilot for channel estimation. To mitigate the interference in the DL, we integrate RSMA in a CF-MaMIMO network using conventional conjugate beamforming (CB) for the private messages, e.g.~\cite{CFMaMIMOVsSmallCells}. We first derive the achievable SE expressions for the common and private streams and then used them to compute an heuristic precoder for the common message as well as to derive a novel successive convex approximation (SCA)-based power control algorithm. Both depend solely on the channel statistics and thus can be used for many coherence intervals. Numerical results are used to show the effectiveness of RSMA in mitigating the effect of pilot contamination.  




\section{System Model}\label{Sysmod}
We consider a cell-free network with $M$ single-antenna APs that are connected via fronthaul links to a central processing unit and serve jointly $K$ single-antenna UEs~\cite{DBS21-book}. The standard time-division-duplexing protocol of cellular Massive MIMO is used~\cite[Sec. 2.3.2]{DBS21-book}, where the $\tau$
available channel uses are employed: $\tau_{p}$ for UL training phase, $\tau_d$ for DL payload transmission and $\tau_u$ for UL payload transmission. Clearly, $\tau\geq\tau_p + \tau_{d} + \tau_{u}$. In this letter, we consider only DL payload transmission and thus we set $\tau_{u}=0$. We denote $g_{mk}\in\mathbb{C}$ the channel between AP $m$ and UE  $k$ and we model it as:
\begin{equation}\label{eq:channel_mk}
g_{mk}=\sqrt{\beta_{mk}}h_{mk},
\end{equation}
where $\beta_{mk}$ represents the large-scale fading (including pathloss and shadow fading) and $h_{mk}\sim\mathcal{CN}(0,1)$ accounts for the small-scale fading. 

We assume that the same pilot sequence is used by all active UEs for channel estimation and that it is transmitted with total power $\rho_{\rm{tr}}$. Hence, the minimum mean square error (MMSE) estimate of $g_{mk}$ is~\cite{luca@pilot}
\begin{equation}\label{eq:channel_gmk_est}
\widehat{g}_{mk}= {\beta}_{mk}{Q}_{m}^{-1}\bigg(\sum_{i=1}^{K}{g}_{mi} + \frac{1}{\sqrt{\rho_{\rm{tr}}}}{z}_{m}\bigg)\end{equation}
where $z_{m}\sim \mathcal{CN}(0,\sigma_{\rm{ul}}^{2})$ is the additive noise at AP $m$ and 
\begin{equation}\label{eq:factor_Qm}
Q_{m}=\sum_{k=1}^{K}{\beta}_{mk} + \frac{\sigma_{\rm{ul}}^{2}}{{\rho_{\rm{tr}}}}.
\end{equation}
The estimates and estimation errors $\widetilde {g}_{mk}= g_{mk} - \widehat{ g}_{mk}$ are independent and distributed as~$\widehat{g}_{mk} \sim \mathcal{CN}\left(0,\gamma_{mk} \right)$ with $\gamma_{mk}=\beta_{mk}^2Q_{m}^{-1}$ and~$\widetilde{g}_{mk} \sim \mathcal{CN}\left(0,\beta_{mk} - \gamma_{mk} \right)$. The interference generated by the pilot-sharing UEs in~\eqref{eq:factor_Qm} is known as pilot contamination~\cite[Sec. 3.3]{massivemimobook}. As in cellular Massive MIMO~\cite[Sec. 4.2.2]{massivemimobook}, it reduces the estimation quality, and makes the estimates correlated with $\mathbb{E}\{\widehat{g}_{mi}\widehat{g}_{mk}^{*}\}=\beta_{mi}Q_{m}^{-1}\beta_{mk}$. This has an important impact beyond channel estimation, since it makes it more difficult to mitigate interference between pilot-sharing UEs.

%
%
%


\subsection{Rate-splitting strategy in the downlink}
We assume that a single-layer rate-splitting (RS) strategy is used in the DL; that is, only one layer of successive interference cancellation (SIC) is applied at the receiver side~\cite{mao2017rate}. We call $W_{k}$ the message intended to UE $k$, which is divided into a common part $W_{c,k}$ and a private part $W_{p,k}$. The common parts of all UEs $\{W_{c,1},\ldots,W_{c,K}\}$ are combined together to form $W_{c}$. Using a common codebook, $W_{c}$ is encoded into a common stream $s_{c}\in\mathbb{C}$ with $\mathbb{E}\{|{s}_{c}|^{2}\}=1$. Notice that the common stream is meant to be decoded by all UEs but not necessarily intended to all of them. The private part $W_{p,k}$ is independently encoded into a private stream $s_{k}\in\mathbb{C}$ with $\mathbb{E}\{|{s}_{k}|^{2}\}=1$, and is meant to be decoded by the corresponding UE only. The signal transmitted by AP $m$ is thus given by
\begin{equation}\label{eq:Tx_APm}
{x}_m=\sqrt{\rho_{T}\eta_{c,m}}{w}_{c,m}s_{c}+\sum_{i=1}^{K}\sqrt{\rho_{T}\eta_{mi}}{w}_{mi}s_{i},
\end{equation}
where ${w}_{c,m}\in\mathbb{C}$ is the precoder of the common stream and ${w}_{mi}\in \mathbb{C}$ is the precoder for the private stream intended to UE $i$. Also, $\rho_{T}$ is the total DL transmit power available at AP $m$ with $\eta_{c,m}$ and $\eta_{mi}$ being the power control coefficients of the common and private streams, respectively. The received signal at UE $k$ is given by
\begin{equation}
 y_{k}=\sum_{m=1}^{M}{g}_{mk}^{*}{x}_m + n_k,
\end{equation}
where $n_{k}\in\mathcal{CN}(0,\sigma^{2})$ is the additive noise. The standard RS decoding scheme is used to form $\widehat{W}_{k}$~\cite{mao2017rate}.

%


\subsection{Spectral efficiency}

Since no pilots are transmitted in the DL, the UE does not know the precoded channels $g_{mk}^{*}w_{c,m}$ and $g_{mk}^{*}w_{mk}$ for $m=1,\ldots M$. Instead, we assume that the UE has access to their expected values $\mathbb{E}\{g_{mk}^{*}w_{c,m}\}$ and $\mathbb{E}\{g_{mk}^{*}w_{mk}\}$. The received signal for the common stream is thus expressed as
\begin{equation}\label{eq:Common_Rx}
y_{c,k}=\mathrm{T}_{c,k}s_{c}+\mathrm{BI}_{c,k}s_{c}+\sum_{i=1}^{K}\mathrm{I}_{ki}s_{i}+n_{k},  
\end{equation}
where 
\begin{align}
\mathrm{T}_{c,k}&=\sqrt{\rho_{T}}\sum_{m=1}^{M}\sqrt{\eta_{c,m}}\mathbb{E}\left\{{g}_{mk}^{*}{w}_{c,m}\right\},\\
\mathrm{BI}_{c,k}&=\sqrt{\rho_{T}}\sum_{m=1}^{M}\sqrt{\eta_{c,m}}{g}_{mk}^{*}{w}_{c,m}-\mathrm{T}_{c,k},\\
\mathrm{I}_{ki}&=\sqrt{\rho_{T}}\sum_{m=1}^{M}\sqrt{\eta_{mi}}{g}_{mk}^{*}{w}_{mi}.  
\end{align}
After cancellation of the common stream, the received signal for the private stream of UE $k$ is expressed as
\begin{align}\label{eq:Private_Rx}
y_{p,k}=\mathrm{T}_{p,k}s_{k}+\mathrm{BI}_{p,k}s_{k}+\mathrm{BI}_{c,k}s_{c}+\sum_{i\neq k}^{K}\mathrm{I}_{ki}s_{i}+n_{k}
\end{align}
where
\begin{align}
&\mathrm{T}_{p,k}=\sqrt{\rho_{T}}\sum_{m=1}^{M}\sqrt{\eta_{mk}}\mathbb{E}\left\{{g}_{mk}^{*}{w}_{mk}\right\},\\
&\mathrm{BI}_{p,k}=\sqrt{\rho_{T}}\sum_{m=1}^{M}\sqrt{\eta_{mk}}{g}_{mk}^{*}{w}_{mk}-\mathrm{T}_{p,k}.
\end{align}
Under the assumption that both $\mathbb{E}\{g_{mk}^{*}w_{c,m}\}$ and $\mathbb{E}\{g_{mk}^{*}w_{mk}\}$ are known, we can compute the following achievable SE for common and private streams at UE $k$ (e.g.,~\cite{luca@pilot,CFMaMIMOVsSmallCells}) 
\begin{align}\label{eq:SE_ck}
\mathrm{SE}_{c,k}&=\frac{\tau_{d}}{\tau}\log_2(1+\mathrm{SINR}_{c,k}),\\ \mathrm{SE}_{p,k}&=\frac{\tau_{d}}{\tau}\log_2(1+\mathrm{SINR}_{p,k}),\label{eq:SE_pk}
\end{align}
where 
\begin{align}\label{eq:sinr_ck}\mathrm{SINR}_{c,k}&=\frac{|\mathrm{T}_{c,k}|^2}{\sum_{i=1}^{K}\mathbb{E}\{|\mathrm{I}_{ki}|^2\}+\mathbb{E}\{|\mathrm{BI}_{c,k}|^{2}\}+\sigma^{2}}\\   
\mathrm{SINR}_{p,k}&=\frac{|\mathrm{T}_{p,k}|^2}{\sum_{i\neq k}^{K}\mathbb{E}\{|\mathrm{I}_{ki}|^2\}+\mathbb{E}\{|\mathrm{BI}_{p,k}|^{2}\}+\mathbb{E}\{|\mathrm{BI}_{c,k}|^{2}\}+\sigma^{2}}
\end{align}
are the effective DL SINRs of the common and private streams at UE $k$, respectively.
As the common stream is decoded by all UEs, we can compute its achievable SE as 
\begin{align}\label{eq:SE_c}
\mathrm{SE}_{c}=\frac{\tau_{d}}{\tau}\log_2(1+\mathrm{SINR}_{c}), 
\end{align}
where $\mathrm{SINR}_{c}=\min_{k}\mathrm{SINR}_{c,k}$. 


\section{Precoder Design}\label{PrecoderDesign}

The achievable SEs in~\eqref{eq:SE_ck} and~\eqref{eq:SE_pk} are very general and can be utilized along
with any precoding scheme for private and common streams. A common and popular choice for the former is CB~\cite{CFMaMIMOVsSmallCells}, defined as $w_{mk}=\widehat{g}_{mk}$, which has low complexity. The precoding scheme for the common stream should be designed to maximize~\eqref{eq:SE_c}. Unfortunately, this is a very challenging problem. To find a low complexity solution, we assume that $\mathbb{E}\{|\mathrm{BI}_{c,k}|^{2}\}$ in~\eqref{eq:sinr_ck} is negligible. This leads to the following sub-optimal design problem:
\begin{equation}\label{eq:Common_Precoder_opt1}
\begin{split}
&\max_{w_{c,m}, \forall m}\;\min_{k}\pi_{k}^{-1}\left|\sum_{m=1}^{M}\mathbb{E}\{g_{mk}^{*}w_{c,m}\}\right|^2\\ &\;\;\;\;{\rm s.t.}\;\;\mathbb{E}\{|w_{c,m}|^2\}=1,\,\forall\,m
\end{split}    
\end{equation}
where $\pi_{k}=\sum_{i=k}^{K}\mathbb{E}\{|\mathrm{I}_{ki}|^2\}+\sigma^{2}$. To further simplify the problem, we assume that $w_{c,m}$ at AP $m$ is obtained as a linear combination of all its estimated channels $\{\widehat{g}_{mi}; i=1,\ldots,K\}$ since it must be decoded by all the UEs. This yields
\begin{equation}\label{eq:Common_Precoder_opt2}
w_{c,m}=\sum_{i=1}^{K}a_{mi}\widehat{g}_{mi}.
\end{equation}
By plugging~\eqref{eq:Common_Precoder_opt2} into~\eqref{eq:Common_Precoder_opt1} and by computing the expectations, we obtain
\begin{equation}\label{eq:Common_Precoder_opt3}
\begin{split}
&\max_{\mathbf{A}}\min_{k}\left(\sum_{m=1}^{M}\sum_{i=1}^{K}a_{mi}\beta_{mi}Q_{m}^{-1}\beta_{mk}\right)^2\\
&\; {\rm s.t.}\; \mathbb{E}\{|\sum_{i=1}^{K}a_{mi}\widehat{g}_{mi}|^2\}=1,\,\forall\,m    
\end{split}    
\end{equation}
where $\mathbf{A}\in \mathbb{R}^{M\times K}$ such that $[\mathbf{A}]_{mi}=a_{mi}$. Introducing auxiliary variable $t$ and vectors $\mathbf{v}=[v_{1},\ldots,v_{M}]^T$ and $\mathbf{u}_{i}=[u_{1i},\ldots,u_{Mi}]^T$ with $u_{mi}=Q_{m}^{-1}\beta_{mi}$, {we transform the objective and constraints of problem~\eqref{eq:Common_Precoder_opt3} into convex form and equivalently transform~\eqref{eq:Common_Precoder_opt3} into the following convex problem}:
\begin{equation}\label{eq:Common_Precoder_opt4}
\begin{split}
 &\max_{\mathbf{A}} \;\;\;t\\
    &\; {\rm s.t.}\; \boldsymbol{\mathbf{u}}_{i}^{T}\mathbf{v}\geq t,\;\;\;\;\;\;\; \forall i,\\
     & \quad  \quad Q_{m}^{-1}v_{m}^{2} \leq 1,\;\;\;\;\;\forall m,\\
    & \quad  \;\; \sum_{i=1}^{K}a_{mi}\beta_{mi}\geq v_{m},\;\; \forall m.
\end{split}    
\end{equation}
Due to its convexity, the solution to~\eqref{eq:Common_Precoder_opt4} can be obtained by using standard convex optimization algorithms, i.e., interior-point method \cite{mishra2021ratesplitting}. In the numerical, we make use of the CVX toolbox in Matlab. Once matrix $\mathbf{A}$ solving~\eqref{eq:Common_Precoder_opt4} is obtained, the common precoder follows from~\eqref{eq:Common_Precoder_opt2}. Notice that~\eqref{eq:Common_Precoder_opt4} depends only on the channel statistics, i.e, large-scale fading. Since they change slowly in time (compared to small-scale fading), the solution to~\eqref{eq:Common_Precoder_opt4} can be used for the design of the common precoder for many coherence intervals.


\section{Power Allocation}\label{PowerAlloc}

We next formulate the novel Max-Min power control algorithm of RSMA in CF-MaMIMO network. We jointly optimize the power control coefficients $\boldsymbol{\eta}=\{\eta_{c,m},\,\eta_{mk}\mid\,\forall m,\,\forall k\}$ and share of the common SE allocated to the UEs $\mathbf{c}=\{C_{1},\ldots,C_{K}\}$, where $C_{k}$ is the share of the common SE allocated to UE $k$. The Max-Min problem is formulated as
\begin{subequations}\label{eq:Max_Min1}
     \begin{align}
          \max_{\boldsymbol{\eta},\mathbf{c}}\;\min_{k}\;\;\;&\textrm{SE}_{p,k}+C_{k}\\
          & C_{1}+\ldots+C_{K} \leq \textrm{SE}_{c,k},\; \forall\,k,\\
          & \eta_{c,m}+\sum_{i=1}^{K}\eta_{mi}\gamma_{mi} \leq 1,\,\forall m,\\
          & \boldsymbol{\eta}\geq \mathbf{0},\\
          & \mathbf{c} \geq \mathbf{0}.
     \end{align}
\end{subequations}
The above problem is non-convex due to the presence of logarithmic and fractional SE expressions $\textrm{SE}_{p,k}$ and $\textrm{SE}_{c,k}$. To solve it, we propose a SCA-based algorithm \cite{mao2018EE} by first defining a vector $\boldsymbol{\zeta}=\sqrt{\boldsymbol{\eta}}$, such that $\zeta_{c,m}=\sqrt{\eta_{c,m}},\,\forall m$, and $\zeta_{mk}=\sqrt{\eta_{mk}},\,\forall m,\,\forall k$. Next we introduce auxiliary vectors $\boldsymbol{\lambda}_{c}=[\lambda_{c,1},\ldots,\lambda_{c,K}]^T$ and $\boldsymbol{\lambda}_{p}=[\lambda_{p,1},\ldots,\lambda_{p,K}]^T$ such that $\lambda_{c,k}$ and $\lambda_{p,k}$ represent the common and private SINRs at UE $k,\,\forall k$. Similarly, we introduce  $\boldsymbol{\alpha}_{c}=[\alpha_{c,1},\ldots,\alpha_{c,K}]^T$ and $\boldsymbol{\alpha}_{p}=[\alpha_{p,1},\ldots,\alpha_{p,K}]^T$ such that $\alpha_{c,k}$ and $\alpha_{p,k}$ represent the common and private SEs at UE $k$. Furthermore, we calculate $\gamma_{c,mk}=\mathbb{E}\{g_{mk}^{*}w_{c,m}\}$, $I_{c,k}^{m}=\mathbb{E}\{|g_{mk}^{*}w_{c,m}|^2\}-|\mathbb{E}\{g_{mk}^{*}w_{c,m}\}|^2$, introduce an auxiliary variable $t$ and equivalently rewrite (\ref{eq:Max_Min1}) as
\begin{subequations}\label{eq:MaxMin_2}
\begin{align}
   \max_{\boldsymbol{\zeta},\boldsymbol{\lambda},\boldsymbol{\alpha},\mathbf{c}}\;\;\;&t\\
   & \alpha_{p,k}+C_{k}\geq t,\; \forall k,\\
   & \alpha_{c,k} \geq \sum_{i=1}^{K}C_{i},\; \forall\,k,\\
   & 1+\lambda_{p,k}-2^{\frac{\tau}{\tau_{d}}\alpha_{p,k}} \geq 0,\forall k,\\
   & 1+\lambda_{c,k}-2^{\frac{\tau}{\tau_{d}}\alpha_{c,k}} \geq 0,\forall k,\\
   &  \zeta_{c,m}^{2}+ \sum_{k=1}^{K}\zeta_{mk} ^{2}\gamma_{mk}\leq 1,\;\forall m, \\
   & (\ref{eq:Max_Min1}\textrm{d}),\,(\ref{eq:Max_Min1}\textrm{e}),\,(\textrm{\ref{eq:lamda_label1}}),\,(\textrm{\ref{eq:lamda_label2}}).
\end{align}
\end{subequations}
Constraints (\ref{eq:lamda_label1}) and (\ref{eq:lamda_label2}) are illustrated at the bottom of this page. The equivalence of (\ref{eq:Max_Min1}) and (\ref{eq:MaxMin_2}) is guaranteed by the fact that (\ref{eq:MaxMin_2}b) is the same as $\min_{k}(\textrm{SE}_{p,k}+C_{k})\geq t$ and it must hold with equality at optimum.
\begin{strip}
\vspace{-0.4cm}
\hrule
\begin{equation}\label{eq:lamda_label1}
   \tag{23h}
   \frac{\rho_{T}(\sum_{m=1}^{M}{\zeta_{c,m}}\gamma_{c,mk})^{2}}{\rho_{T}\sum_{i=1}^{K}\sum_{m=1}^{M}\zeta_{mi}^{2}{\gamma_{mi}}\beta_{mk}+\rho_{T}\sum_{i=1}^{K}(\sum_{m=1}^{M}\zeta_{mi}{\gamma_{mi}}\frac{\beta_{mk}}{\beta_{mi}})^{2}+\rho_{T}\sum_{m=1}^{M}\zeta_{c,m}^{2}I_{c,k}^{m}+\sigma^2}\geq  \lambda_{c,k},\,\forall k
\end{equation}
\begin{equation}\label{eq:lamda_label2}
    \tag{23i}
    \frac{\rho_{T}(\sum_{m=1}^{M}{\zeta_{mk}}{\gamma_{mk}})^{2}}{\rho_{T}\sum_{i=1}^{K}\sum_{m=1}^{M}\zeta_{mi}^{2}{\gamma_{mi}}\beta_{mk}+\rho_{T}\sum_{i\neq k}^{K}(\sum_{m=1}^{M}\zeta_{mi}{\gamma_{mi}}\frac{\beta_{mk}}{\beta_{mi}})^{2}+\rho_{T}\sum_{m=1}^{M}\zeta_{c,m}^{2}I_{c,k}^{m}+\sigma^2}\geq \lambda_{p,k},\,\forall k
\end{equation}
\end{strip}
In problem (\ref{eq:MaxMin_2}), the constraints (\ref{eq:lamda_label1}) and (\ref{eq:lamda_label2}) are non convex. Both can be put into a equivalent convex form by declaring the auxiliary vectors $\boldsymbol{\delta}_{p}=[\delta_{p,1},\ldots,\delta_{p,k}]^T$, $\boldsymbol{\delta}_{c}=[\delta_{c,1},\ldots,\delta_{c,k}]^T$,  $\boldsymbol{\chi}_{c}=[\chi_{c,1},\ldots,\chi_{c,k}]^T$, $\boldsymbol{\chi}_{p}=[\chi_{p,1},\ldots,\chi_{p,k}]^T$, $\boldsymbol{\nu}=[\nu_{1},\ldots,\nu_{M}]^T$ and  $\mathbf{Z}=\{z_{ki}\mid \forall k,i\}$ such that
\begin{subequations}\label{eq:MaxMin_3}
\begin{align}
   & \frac{\delta_{p,k}^{2}}{\chi_{p,k}} \geq \lambda_{p,k},\;\forall k,\\
   & \frac{\delta_{c,k}^{2}}{\chi_{c,k}} \geq \lambda_{c,k},\;\forall k,\\
   & \sum_{m=1}^{M}\zeta_{mk}{\gamma_{mk}} \geq \delta_{p,k},\;\forall k,\\
   & \sum_{m=1}^{M}\zeta_{c,m}\gamma_{c,mk} \geq \delta_{c,k},\;\forall k,\\
   & \chi_{p,k} \geq \sum_{m=1}^{M}\beta_{mk}\nu_{m}^{2}+\sum_{i\neq k}^{K}z_{ki}^{2}+\sum_{m=1}^{M}\zeta_{c,m}^{2}I_{c,k}^{m}+\frac{\sigma^2}{\rho_{T}},\;\forall k,\\
   & \chi_{c,k} \geq \sum_{m=1}^{M}\beta_{mk}\nu_{m}^{2}+\sum_{i=1}^{K}z_{ki}^{2}+\sum_{m=1}^{M}\zeta_{c,m}^{2}I_{c,k}^{m}+\frac{\sigma^2}{\rho_{T}},\;\forall k,\\
   & \nu_{m}^{2} \geq \sum_{i=1}^{K}\zeta_{mi}^{2}\gamma_{mi},\;\forall m,\\
   & z_{ki} \geq \sum_{m=1}^{M}{\gamma_{mi}}{\zeta_{mi}}\frac{\beta_{mk}}{\beta_{mi}},\;\forall k,i,\\
   & {0}\leq \nu_{m}^{2}+\zeta_{c,m}^{2}\leq 1,\;\forall m,\\
   &  \nu_{m} \geq 0,\;\forall m.
\end{align}
\end{subequations}
The constraints in (\ref{eq:MaxMin_3}) are all convex with the exception of (\ref{eq:MaxMin_3}a) and (\ref{eq:MaxMin_3}b). By applying first order Taylor approximation on the left-hand-side of both, we obtain:
\begin{align*}
    \frac{\delta_{c,k}^{2}}{\chi_{c,k}}\geq & \Big(\frac{2\delta_{c,k}^{[n]}}{\chi_{c,k}^{[n]}}\delta_{c,k}-\frac{(\delta_{c,k}^{[n]})^{2}}{(\chi_{c,k}^{[n]})^{2}}\chi_{c,k}\Big)\triangleq \Psi_{c,k}^{[n]}(\delta_{c,k},\chi_{c,k}),\;\forall k,\\
    \frac{\delta_{p,k}^{2}}{\chi_{p,k}}\geq & \Big(\frac{2\delta_{p,k}^{[n]}}{\chi_{p,k}^{[n]}}\delta_{p,k}-\frac{(\delta_{p,k}^{[n]})^{2}}{(\chi_{p,k}^{[n]})^{2}}\chi_{p,k}\Big)\triangleq \Psi_{p,k}^{[n]}(\delta_{p,k},\chi_{p,k}),\;\forall k.
\end{align*}
For a given initialization, the Max-Min power control problem at iteration $[n]$ of the proposed RSMA-assisted CF-MaMIMO can thus be formulated as
\begin{equation}\label{eq:MaxMin_5}
\begin{split}
   \max_{\substack{t,\boldsymbol{\zeta},\mathbf{c},\boldsymbol{\delta}_{c},\boldsymbol{\delta}_{p},\boldsymbol{\chi}_{c},\boldsymbol{\chi}_{p}\\\boldsymbol{\nu},\mathbf{Z},\boldsymbol{\alpha}_{c},\boldsymbol{\alpha}_{p},\boldsymbol{\lambda}_{c},\boldsymbol{\lambda}_{p}}}\;\;&t\\
   &  \Psi_{c,k}^{[n]}(\delta_{c,k},\chi_{c,k})\geq \lambda_{c,k},\;\forall k,\\
   &  \Psi_{p,k}^{[n]}(\delta_{p,k},\chi_{p,k})\geq \lambda_{p,k},\;\forall k,\\
   & (\ref{eq:Max_Min1}\textrm{d}),\,(\ref{eq:Max_Min1}\textrm{e}),\,(\ref{eq:MaxMin_2}\textrm{b})-(\ref{eq:MaxMin_2}\textrm{e}),\\
   &(\ref{eq:MaxMin_3}\textrm{c})-(\ref{eq:MaxMin_3}\textrm{j}).
\end{split}
\end{equation}
Problem (\ref{eq:MaxMin_5}) is convex and like~\eqref{eq:Common_Precoder_opt4} can be solved using the CVX toolbox of Matlab. The SCA-based algorithm to solve problem (\ref{eq:MaxMin_5}) is outlined in Algorithm \ref{alg:Max_MinA}. The power control coefficients are initialized as $\eta_{c,m}=\varrho\rho_{T}$ and $\eta_{mk}=\varrho\rho_{T}/(\sum_{k=1}^{K}\gamma_{mk}),\,\forall k$, $\varrho\in[0,1]$. The initial values $\boldsymbol{\delta}_{p}^{[0]}$,  $\boldsymbol{\delta}_{c}^{[0]}$, $\boldsymbol{\chi}_{p}^{[0]}$ and $\boldsymbol{\chi}_{c}^{[0]}$ are obtained by replacing the inequalities of (\ref{eq:MaxMin_3}c), (\ref{eq:MaxMin_3}d), (\ref{eq:MaxMin_3}e) and (\ref{eq:MaxMin_3}f) with equalities, respectively. In each iteration, problem (\ref{eq:MaxMin_5}) is solved and $\boldsymbol{\delta}_{c}^{[n]}$, $\boldsymbol{\delta}_{p}^{[n]}$, $\boldsymbol{\chi}_{c}^{[n]}$ and $\boldsymbol{\chi}_{p}^{[n]}$  are updated. At the output of the $n^{th}$ iteration, $t^{[n]}$ is the maximized minimum SE. The tolerance of the algorithm is $\epsilon$.
\par As (\ref{eq:MaxMin_3}a) and (\ref{eq:MaxMin_3}b) are relaxed by the first-order lower bounds $ \Psi_{p,k}^{[n]}(\delta_{p,k},\chi_{p,k})$ and $\Psi_{c,k}^{[n]}(\delta_{c,k},\chi_{c,k})$ respectively, the solution of problem (\ref{eq:MaxMin_5}) at iteration $[n]$ is also a feasible solution at iteration $[n+1]$. As a result, the optimized $t$ is non-decreasing as iteration increases, i.e., $t^{[n+1]} \geq t^{[n]}$ will always hold. Furthermore, the objective of problem (\ref{eq:MaxMin_5}) is bounded by the DL transmit power constraint  (\ref{eq:Max_Min1}c) and thus is guaranteed to converge. However, since the constraints (\ref{eq:MaxMin_3}a) and (\ref{eq:MaxMin_3}b) are linearly approximated, {convergence to the global optimal solution with RS is not guaranteed}.
\begin{algorithm}
\caption{MaxMin: SCA Algorithm}\label{alg:Max_MinA}
\begin{algorithmic}[1]
\State $\mathbf{Initialize}\:\: n\leftarrow 0,\: t^{[n]}\leftarrow 0, \varrho\leftarrow 0.5,\,\boldsymbol{\delta}_{c}^{[n]}, \boldsymbol{\delta}_{p}^{[n]},  \boldsymbol{\chi}_{p}^{[n]}, \boldsymbol{\chi}_{c}^{[n]}$
\State $\mathbf{Iterate}$
\State $\;\;n\leftarrow n+1;$
\State $\;\;Solve\;(\ref{eq:MaxMin_5})\;\textrm{using}$
\item[]$\;\;\;\;\boldsymbol{\delta}_{p}^{[n-1]}, \boldsymbol{\delta}_{c}^{[n-1]}, \boldsymbol{\chi}_{p}^{[n-1]}, \boldsymbol{\chi}_{c}^{[n-1]} \textrm{and}\;\textrm{denote}\;\textrm{optimal}\;\textrm{values}$
\item[]$\;\;\;\;\;\textrm{of}\;\;\;t, \boldsymbol{\delta}_{c},  \boldsymbol{\delta}_{p},\,\boldsymbol{\chi}_{p},\,\boldsymbol{\chi}_{c} \;\;\;\textrm{as}\;\;\; t^{*},\,\boldsymbol{\delta}_{c}^{*},\, \,\boldsymbol{\delta}_{p}^{*},\,\boldsymbol{\chi}_{p}^{*},\,\boldsymbol{\chi}_{c}^{*}.$
\State $\mathbf{Update}\;t^{[n]}\leftarrow t^{*},\,\boldsymbol{\delta}_{c}^{[n]}\leftarrow\boldsymbol{\delta}_{c}^{*},\,\boldsymbol{\delta}_{p}^{[n]}\leftarrow\boldsymbol{\delta}_{p}^{*},\,\boldsymbol{\chi}_{p}^{[n]}\leftarrow \boldsymbol{\chi}_{p}^{*},$
\item[]$\;\;\;\;\;\;\;\;\;\;\;\;\;\;\;\boldsymbol{\chi}_{c}^{[n]}\leftarrow \boldsymbol{\chi}_{c}^{*}$
\State $\mathbf{Until}\;\;|t^{[n]}-t^{[n-1]}|< \epsilon $
\end{algorithmic}
\end{algorithm}


\section{Numerical Results}\label{NumRes}
Numerical simulations are now used to assess the performance of the proposed RSMA scheme.
Two different network topologies are considered: 1) a rectangular area of side $250\,\textrm{m}$ where APs and UEs are randomly distributed (Rect); and 2) a circular area of radius  $r=125\,\textrm{m}$, where APs are uniformly distributed in the circumference and UEs are randomly distributed inside the circle (Circ). The large scale fading coefficient $\beta_{mk}$ between AP $m$ and UE $k$ is modelled in dB as $\beta_{mk}[\mathrm{dB}]=\mathrm{P}_{mk}+S_{mk}$ where $\mathrm{P}_{mk}$ is exactly in the same form of~\cite[Eq. (52)]{CFMaMIMOVsSmallCells} and $S_{mk}$ represents the independent shadow fading with variance equal to~$16$. The simulation  parameters are reported in Table~\ref{tab:Contri_table}.

\begin{table}
	\caption{Simulation parameters}
	    \label{tab:Contri_table}\centering
	\begin{tabular}{l l}
		\toprule[0.4mm]
		\textbf{Parameter} & \textbf{Value}\\
   {Carrier frequency} & {$1.9\,\textrm{GHz}$}\\
   {AP and UE heights for~\cite[Eq. (52)]{CFMaMIMOVsSmallCells}} & $h_{\textrm{AP}} = 15\,\textrm{m}$, $h_{\textrm{u}} = 1.65\,\textrm{m}$\\
   {Reference distances for~\cite[Eq. (52)]{CFMaMIMOVsSmallCells}} & $d_{1} = 50\,\textrm{m}$, $d_{0}=10\,\textrm{m}$\\
   TDD parameters (samples) & {$\tau = 200,\,\tau_{p}=10,\,\tau_{d}= 190$}\\
   Total transmit powers &  $\rho_{\rm{tr}} = 10\,\textrm{dBm}$, $\rho_{T}= 20\,\textrm{dBm}$\\
   Noise powers & $\sigma_{\rm{ul}}^{2}=\sigma^{2} = -94\,\textrm{dBm}$\\
		\bottomrule[0.4mm]
	\end{tabular}
\end{table}

\begin{figure}[t!]\vspace{-0.4cm}
	\centering 
	\begin{overpic}[width=0.8\columnwidth,tics=10]{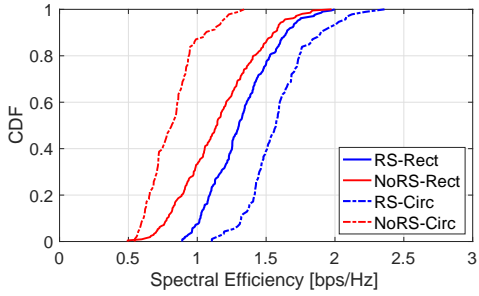}
\end{overpic} 
    \caption{CDF plot of SE per UE for $M=100$ and $K=10$ in the two considered network layouts.}
    \label{fig:CDF_plot}\vspace{-0.4cm}
\end{figure}

\begin{figure}
    \centering
    \subfloat[$M=100$]{\includegraphics[width=0.8\columnwidth,tics=10]{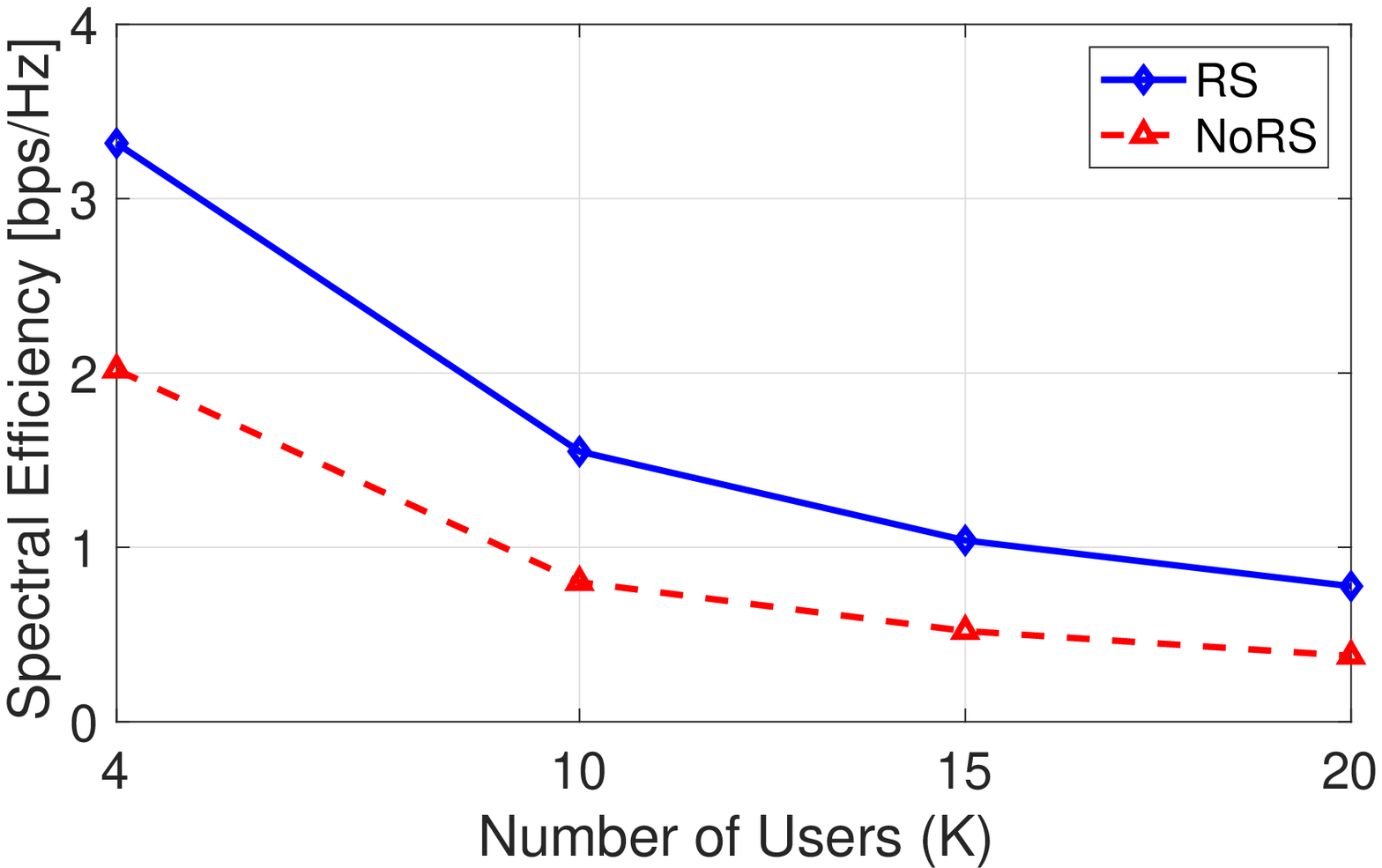}}\\\vspace{-0.4cm}
    \subfloat[$K=10$]{\includegraphics[width=0.8\columnwidth,tics=10]{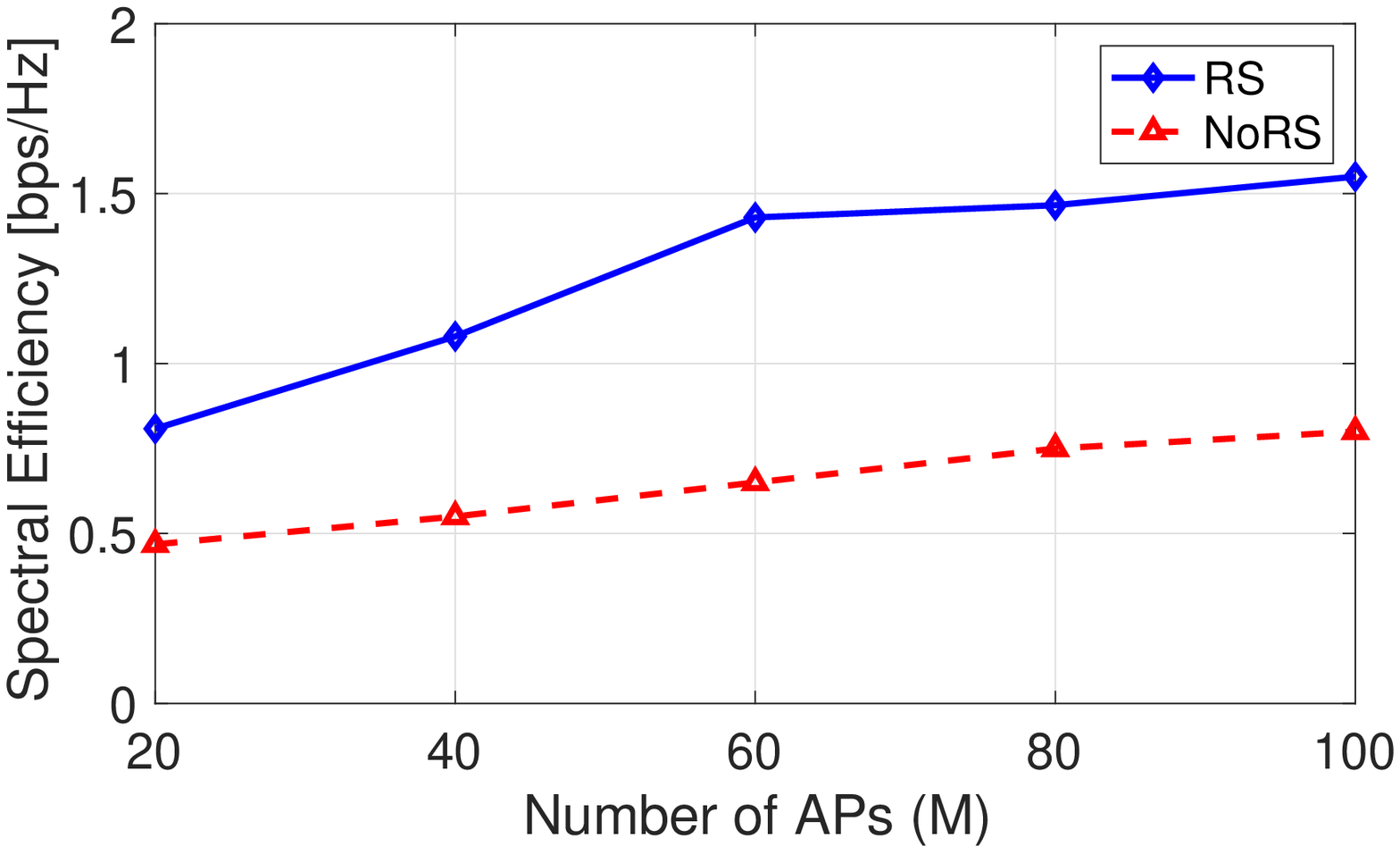}}
    \caption{Average SE per UE  with a circular topology.}
    \label{fig:SR_plot}\vspace{-0.4cm}
\end{figure}
\par Fig.~\ref{fig:CDF_plot} illustrates the CDF plot of the SE per UE with both topologies. We see that in both cases RS provides substantial gains compared to the case where no RS\footnote{For simplicity, a conventional linearly precoded transmission scheme is referred to as NoRS \cite{massiveMIMO@hardware}. {Algorithm 2 in \cite{CFMaMIMOVsSmallCells} used here to obtain the globally optimal result without RS, employs CB and bisection algorithm}.} is used~\cite[Alg.~2]{CFMaMIMOVsSmallCells}. This is because RS allows to allocate a portion of the total DL power to the common message when imperfect CSI is available and thus handle the interference more effectively at the UE side. We see that the gain is larger with a circular topology. This is due to the following two factors. Firstly, in a circular topology the UEs experience comparable path losses and thus the effect of pilot contamination is relatively uniform across all APs. In contrast, the effect of pilot contamination varies significantly across APs in the rectangular topology due to the random distribution of APs. {This variation aids the Max-Min power control strategy, which effectively acts as the AP selection strategy for a UE, in selecting APs at which the pilot of the corresponding UE is least contaminated. Therefore, the effect of pilot contamination is less severe in rectangular topology}. Secondly, since the common message needs to be decoded at all UEs, comparable path losses aids the common precoder design as well as the maximization of the common SE, thereby increasing the performance of RSMA. 

With circular topology, in Fig.~\ref{fig:SR_plot}(a) we observe that as $K$ increases from $4$ to $20$, the CSI quality deteriorates further and consequently the average SE per UE decreases with and without RS. Furthermore, as $K$ increases, the gain with RS decreases because maximizing the common SE becomes more difficult since the common message has to be decoded by all UEs. However, the gain is still significant. Fig.~\ref{fig:SR_plot}(b) illustrates that RS outperforms NoRS in average SE per UE performance even with finite $M$. Moreover, as $M$ increases, the gain with RS increases because SIC of the common message is aided by channel hardening with increasing $M$ and RS better manages the interference. Same behaviours can be observed for the rectangular topology, but are not shown for space limitation.


\section{Conclusions}

In this letter, we propose a novel system model of RSMA in time-division-multiplexing CF-MaMIMO and employed RSMA as a pilot contamination mitigation strategy in the DL of mMTC with random access. By devising a heuristic common precoder design and formulating a novel Max-Min power control algorithm, we illustrated the achievable SE performance of RSMA in CF-MaMIMO with different AP topologies. Numerical results showed that with single pilot used
for UL channel estimation, the SE achieved with RSMA is significantly higher than that of a conventional CF-MaMIMO scheme.

\ifCLASSOPTIONcaptionsoff
  \newpage
\fi

\appendices

\bibliographystyle{IEEEtran}
\bibliography{refer}

\begin{thebibliography}{10}
\providecommand{\url}[1]{#1}
\csname url@samestyle\endcsname
\providecommand{\newblock}{\relax}
\providecommand{\bibinfo}[2]{#2}
\providecommand{\BIBentrySTDinterwordspacing}{\spaceskip=0pt\relax}
\providecommand{\BIBentryALTinterwordstretchfactor}{4}
\providecommand{\BIBentryALTinterwordspacing}{\spaceskip=\fontdimen2\font plus
\BIBentryALTinterwordstretchfactor\fontdimen3\font minus
  \fontdimen4\font\relax}
\providecommand{\BIBforeignlanguage}[2]{{%
\expandafter\ifx\csname l@#1\endcsname\relax
\typeout{** WARNING: IEEEtran.bst: No hyphenation pattern has been}%
\typeout{** loaded for the language `#1'. Using the pattern for}%
\typeout{** the default language instead.}%
\else
\language=\csname l@#1\endcsname
\fi
#2}}
\providecommand{\BIBdecl}{\relax}
\BIBdecl

\bibitem{Schober_mMTC}
X.~Chen, D.~W.~K. Ng, W.~Yu, E.~G. Larsson, N.~Al-Dhahir, and R.~Schober,
  ``Massive access for 5{G} and beyond,'' \emph{IEEE J. Sel. Areas Commun.},
  vol.~39, no.~3, pp. 615--637, 2021.

\bibitem{erykMTC@2017}
E.~Dutkiewicz, X.~Costa-Perez, I.~Z. Kovacs, and M.~Mueck, ``Massive
  machine-type communications,'' \emph{IEEE Network}, vol.~31, no.~6, pp. 6--7,
  2017.

\bibitem{luca@pilot}
A.~A. Polegre, L.~Sanguinetti, and A.~G. Armada, ``Pilot decontamination
  processing in cell-free massive {MIMO},'' \emph{{IEEE} Commun. Lett.}, pp.
  1--1, 2021.

\bibitem{carvalho@randomaccess}
J.~H. Sørensen, E.~de~Carvalho, and P.~Popovski, ``Massive {MIMO} for crowd
  scenarios: A solution based on random access,'' in \emph{2014 IEEE Globecom
  Workshops (GC Wkshps)}, 2014, pp. 352--357.

\bibitem{CFMaMIMOVsSmallCells}
H.~Q. {Ngo}, A.~{Ashikhmin}, H.~{Yang}, E.~G. {Larsson}, and T.~L. {Marzetta},
  ``Cell-free massive {MIMO} versus small cells,'' \emph{IEEE Trans. Wireless
  Commun.}, vol.~16, no.~3, pp. 1834--1850, 2017.

\bibitem{Yin@singlepilot}
H.~Yin, D.~Gesbert, and L.~Cottatellucci, ``Dealing with interference in
  distributed large-scale {MIMO} systems: A statistical approach,'' \emph{IEEE
  J. Sel. Topics in Signal Process.}, vol.~8, no.~5, pp. 942--953, 2014.

\bibitem{massivemimobook}
E.~Bj\"{o}rnson, J.~Hoydis, and L.~Sanguinetti, ``Massive {MIMO} networks:
  {Spectral}, energy, and hardware efficiency,'' \emph{Foundations and
  Trends{\textregistered} in Signal Processing}, vol.~11, no. 3-4, pp.
  154--655, 2017.

\bibitem{mao2017rate}
Y.~Mao, B.~Clerckx, and V.~O.~K. Li, ``Rate-splitting multiple access for
  downlink communication systems: bridging, generalizing, and outperforming
  {SDMA} and {NOMA},'' \emph{{EURASIP} J. Wireless Commun. Netw.}, vol. 2018,
  no.~1, p. 133, May 2018.

\bibitem{Onur@6G}
O.~Dizdar, Y.~Mao, W.~Han, and B.~Clerckx, ``Rate-splitting multiple access: A
  new frontier for the {PHY} layer of 6{G},'' in \emph{2020 IEEE 92nd Vehicular
  Technology Conference (VTC2020-Fall)}, 2020, pp. 1--7.

\bibitem{mishra2021ratesplitting}
A.~Mishra, Y.~Mao, O.~Dizdar, and B.~Clerckx, ``Rate-splitting multiple access
  for downlink multiuser {MIMO}: Precoder optimization and {PHY}-layer
  design,'' \emph{IEEE Trans. Commun.}, pp. 1--1, 2021.

\bibitem{mao2018EE}
Y.~Mao, B.~Clerckx, and V.~O.~K. Li, ``Energy efficiency of rate-splitting
  multiple access, and performance benefits over {SDMA} and {NOMA},'' in
  \emph{Proc. {IEEE} Int. Symp. Wireless Commun. Syst. (ISWCS)}, Aug. 2018, pp.
  1--5.

\bibitem{massiveMIMO@hardware}
A.~Papazafeiropoulos, B.~Clerckx, and T.~Ratnarajah, ``Rate-splitting to
  mitigate residual transceiver hardware impairments in massive {MIMO}
  systems,'' \emph{{IEEE} Trans. Veh. Technol.}, vol.~66, no.~9, pp.
  8196--8211, 2017.

\bibitem{DBS21-book}
\BIBentryALTinterwordspacing
{\"O}.~T. Demir, E.~Bj{\"o}rnson, and L.~Sanguinetti, ``{Foundations of
  User-Centric Cell-Free Massive MIMO},'' \emph{Foundations and Trends in
  Signal Processing}, vol.~14, no. 3-4, pp. 162--472, 2021. [Online].
  Available: \url{http://dx.doi.org/10.1561/2000000109}
\BIBentrySTDinterwordspacing

\bibitem{Lamare@2020}
R.~B. Di~Renna and R.~C. de~Lamare, ``Iterative list detection and decoding for
  massive machine-type communications,'' \emph{IEEE Transactions on
  Communications}, vol.~68, no.~10, pp. 6276--6288, 2020.

\end{thebibliography}

\end{document}